\documentclass[11pt]{article}
\usepackage{cite}
\usepackage{amsmath}
\usepackage{amssymb}

\newcommand{\versionsize}{100}
\newcommand{\maybe}[3]{\ifnum \versionsize < #1 #2\else #3\fi}

\maybe{100}{

\setlength{\textheight}{8.5in}
\setlength{\textwidth}{6.5in}
\setlength{\topmargin}{0pt}
\setlength{\evensidemargin}{-0.0 in}
\setlength{\oddsidemargin}{-0.0 in}
\setlength{\headsep}{10pt}
\setlength{\parskip}{\medskipamount}
\setlength{\itemsep}{-.6ex}
}{
\addtolength{\textheight}{1.4in}
\addtolength{\textwidth}{1.6in}
\addtolength{\topmargin}{-0.7in}
\addtolength{\evensidemargin}{-0.8in}
\addtolength{\oddsidemargin}{-0.8in}
}

\newcommand{\alert}[1]{\typeout{ALERT: #1}\textbf{[[[ #1 ]]]}}

\newtheorem{theorem}{Theorem}
\newtheorem{lemma}[theorem]{Lemma}
\newtheorem{corollary}[theorem]{Corollary}
\newenvironment{proof}{\noindent\par{\bf Proof: }}{\nopagebreak\rule{1 ex}{0.8 em}\medskip}

\newcommand{\newloglike}[2]{\newcommand{#1}{\mathop{\rm #2}\nolimits}}
\newloglike{\E}{E}

\newloglike{\wt}{weight}
\newloglike{\val}{value}
\newloglike{\select}{select}

\newcommand{\pred}{\rightarrow}

\newcommand{\lt}{<}

\newloglike{\treeroot}{root}
\newloglike{\opcostvalue}{opcost}

\newcommand{\opcost}[1]{{\sc Op\-cost}#1}
\newcommand{\lropcost}[1]{{\sc LR-}\opcost{#1}}
\newcommand{\unweighted}[1]{{\sc Un\-weighted-}#1}
\newcommand{\overlapping}[1]{{\sc Over\-lapping-}#1}
\newcommand{\weighted}[1]{{\sc Weighted-}#1}
\newcommand{\light}[1]{{\sc Light-}#1}
\newcommand{\heavy}[1]{{\sc Heavy-}#1}
\newcommand{\oc}{OC}
\newcommand{\lr}{LR}

\newcommand{\etal}[1]{{\it et al.\/}}
\newcommand{\buzz}[1]{\emph{#1}}
\newloglike{\tw}{tw}

\newcommand{\objs}{{\cal O}}
\newcommand{\bids}{{\cal B}}
\newcommand{\allbids}{{\cal A}}

\newcommand{\suppress}[1]{}

\newcommand{\R}{\mathbb{R}}

\title{Opportunity Cost Algorithms for Combinatorial Auctions}
\author{
Karhan Akcoglu\thanks{
Department of Computer Science, Yale University,
New Haven, CT 06520-8285, USA.
Email: {\tt karhan.akcoglu@yale.edu}. 
Supported in part by NSF Grant CCR-9896165.}
\and
James Aspnes\thanks{
Department of Computer Science, Yale University,
New Haven, CT 06520-8285, USA.
Email: {\tt aspnes-james@cs.yale.edu}.
Supported in part by NSF grant CCR-9820888.}
\and
Bhaskar DasGupta\thanks{
Department of Computer Science,
Rutgers University,
Camden, NJ 08102, USA.
Email: {\tt bhaskar@crab.rutgers.edu}.
Supported in part by NSF Grant CCR-9800086.
}
\and
Ming-Yang Kao\thanks{Department of Computer Science, Yale University,
New Haven, CT 06520-8285, USA.  Email: {\tt kao-ming-yang@cs.yale.edu}. 
Supported in part by NSF Grant CCR-9531028.}
}

\begin{document}

\maketitle

\begin{abstract}
Two general algorithms based on opportunity costs are given for approximating
a revenue-maximizing set of bids an auctioneer should accept, in a
\buzz{combinatorial auction} in which each bidder offers a price for
some subset of the available goods and the auctioneer can only
accept non-intersecting bids.  Since this problem is difficult
even to approximate in general,
the algorithms are most useful when the bids
are restricted
to be connected node subsets of an underlying \buzz{object graph} that
represents which objects are relevant to each other.  The approximation
ratios of the algorithms depend on structural properties of this graph
and are small constants for many interesting families of object graphs.
The running times of the algorithms are linear in the size of the
\buzz{bid graph}, which describes the conflicts between bids.  Extensions
of the algorithms allow for efficient processing of additional constraints,
such as budget constraints that associate bids with particular bidders
and limit how many bids from a particular bidder can be accepted.
\end{abstract}

\section{Introduction}
\label{section-intro}

Auctions are arguably the simplest and most popular means of price
determination for multilateral trading without intermediary market
makers \cite{HP93,McA87,Wil92,clearwater}.  This paper considers the
setting where there are (1) a group of competing {\it bidders} who bid
to possess the auction {\it objects} and (2) an {\it auctioneer} who
determines which bidders win which objects.

For the case of allocating a single object to one of many bidders,
there is a wealth of literature on the following four widely used forms
of auction \cite{HP93,McA87,MW82}.  In an \emph{English auction} or
\emph{ascending bid auction}, the price of an object is successively
raised until only one bidder remains and wins the object.  In a
\emph{Dutch auction}, which is the converse of an English auction, an
initial high price is subsequently lowered until a bidder accepts the
current price.  In a \emph{first-price sealed-bid auction}, potential
buyers submit sealed bids for an object. The highest bidder is awarded
the object and pays the amount of her bid.  In a \emph{second-price
sealed-bid auction}, the highest bidder wins the object but pays a
price equal to the second-highest bid. In all these forms of auction,
the auctioneer can determine the winning bid in time linear in the
number of bids in a straightforward manner.

For the case of allocating multiple objects to multiple bidders\cite{KR96,Hausch86,Palfrey80,Gale90,kaoqt.auction.sjp,ckao.bidding.scp}, 
{\it
combinatorial auctions} are perhaps the most important form of
auctions
in the Internet Age, where bidders are increasingly software agents.
Oftentimes a bid by an agent is a subset of the auction objects, and
the agent needs the entire subset to complete a task.  Different bids
may share the same object, but the winning bids must not share any
object \cite{MM1996}. Combinatorial auctions were first proposed by
Rassenti \etal{Smith, and Bulfin} \cite{RSB1982} as one-round mechanisms
for airport time slot allocation.  Banks \etal{Ledyard, and Porter}
\cite{BLP1989}, DeMartini \etal{Kwasnica, Ledyard, and Porter}
\cite{DKLP1999}, and Parkes and Ungar \cite{PU2000} formulated
multiple-round mechanisms.  It is in general $NP$-hard for the
auctioneer to determine a set of winning bids of a combinatorial
auction which maximizes the revenue of the auction.  To address this
computational difficulty, Rothkopf \etal{Peke\v{c}, and Harstad}
\cite{RPH1998} placed constraints on permissible bids.  
Lehmann \etal{O'Callaghan, and Shoham}
\cite{LOS1999} and Fujishima \etal{Leyton-Brown, and Shoham} \cite{FLS99}
considered approximation algorithms. Sandholm and Suri \cite{SS2000}
designed \buzz{anytime} algorithms, which return a sequence of
monotonically improving solutions that eventually converges to optimal.

In this paper, we propose a general framework to exploit topological
structures of the bids to determine the winning bids with a provably
good approximation ratio in linear time.  The following discussion
uses the sale of a car
as a light-hearted example to explain our computational
problems and key concepts.

Imagine that we are in the business of auctioning used cars.  If we
insist on selling each car as a unit, we can sell each car to the
highest bidder.  If we are willing to sell parts of the car, we can
still sell each part to the highest bidder.  But suppose that some
bidders are only interested in buying several parts at once: Alice may
not want to buy a tire unless she can get the wheel that goes with it,
while Bob might only be interested in both rear wheels and the axle
between them.  How do we decide which of a set of conflicting bids to
accept?

We will assume that our only goal is to maximize our total
revenue. Then we can express this problem as a simple combinatorial
optimization problem.  We have some universe $\objs$ of objects, and
our buyers supply us with a set $\allbids$ of bids.  The $i$-th bid
consists of a subset $A_i$ of $\objs$ and a price $p_i$ that the buyer
is willing to pay for all of the objects in $S_i$.  We would like to
choose a collection of bids $\bids \subseteq \allbids$ that
yields the best possible total price while being
\buzz{consistent}, in the sense that no two sets $A_i$ and $A_j$ in
$\bids$ overlap.

As the auctioneer, we can construct a \buzz{bid graph} $G$ whose nodes are
the
bids and which has an edge between any two bids that share an object.
Then, a set of consistent bids is simply an independent set in $G$, 
i.e., a set of nodes no two of which are connected by an
edge.  Each node is given a weight equal to the value of the bid it
represents.

Sadly, this means that
the problem of finding the most valuable consistent set of bids
is a thinly-disguised version of the
maximum weight independent set problem, which is not only $NP$-hard, but
cannot be approximated to within a ratio 
$O(n^{1-\epsilon})$ 
for an $n$-node graph
unless $P = NP$ \cite{hastad1999.clique}.\footnote{
The fact that the bid graph is defined by the
intersections of a collection of sets does not by itself help; any graph can be
defined in this way.}
Even for the simplest case when all node weights are one, 
the maximum weight independent set problem is $NP$-hard even when 
every vertex
has degree at most $d$ for any $d\geq 3$, and in fact cannot be 
approximated within a ratio of $d^{\varepsilon}$ for some 
$\varepsilon>0$ unless $P=NP$~\cite{AlonFWZ1995}. The best known algorithm 
(for arbitrary $d$) achieves an approximation within a factor of 
$O(d/\log\log d)$~\cite{HalldorssonR1994}. 
As a result, it seems hopeless if we
model our combinatorial auction problem as an independent set
problem unless we exploit the topological structure of the 
underlying bid graph. 

Using ideas from the interval selection algorithm of
Berman and DasGupta
\cite{BermanDasGupta}, we describe in Section \ref{section-noBudgetConstraints}
a linear-time improvement of the greedy algorithm, called the
\emph{opportunity cost algorithm}, for approximating
maximum weight independent sets in ordered graphs.\footnote{These are
graphs in which the nodes have been assigned an order; as we will
see in Section \ref{section-ordering}, the choice of order for a given
bid graph can have a large effect on how good an approximation we can
get.}  
We then describe a similar algorithm called the 
\buzz{local ratio opportunity cost algorithm}, based on ideas 
from the resource allocation algorithms of 
Bar-Noy \etal~\cite{BarNoyetal2000}.
Both algorithms produce the same output, but the first has a more
iterative structure and is easier to implement while the second has a
more recursive structure and is easier to analyze.

These opportunity cost algorithms distinguish themselves from the
straightforward greedy algorithm by taking into account the cost of excluding
previously considered neighbors of a chosen node.  Since this
accounting requires propagating information only between neighbors, it
increases the running time by at most a small constant factor, and yet
in many cases produces a great improvement in the approximation ratio.
The quality of the approximation depends on the local structure of the
ordered input graph $G$.  For each node $v$ in $G$, we examine
all of its successors (adjacent nodes that appear later in the
ordering).  The maximum size of any independent set among $v$ and its
successors is called the \buzz{directed local independence number at
$v$}; we will write it as $\beta(v)$.  The maximum value of
$\beta(v)$ over all nodes in the graph will be written as
$\beta(G)$,\footnote{Or simply $\beta$ when $G$ is clear from the
context.}
and is the {\it directed local independence number} of $G$.  Our algorithms
approximate a maximum weight independent set to within a factor of
$\beta$.  By comparison, the greedy algorithm approximates a
maximum weight independent set within a ratio of the maximum size of
any independent subset of both the predecessors and the successors of
any node, which in general can be much larger than $\beta$ (see
Section
\ref{section-noBudgetConstraints}).

These new approximation results are useful only if we can exhibit interesting
classes of graphs for which $\beta$ is small.  Graphs with $\beta=1$
have been extensively studied in the graph theory literature; these
are known as \buzz{chordal graphs}, and are precisely those graphs
that can be represented as intersection graphs of subtrees of a forest, a class
that includes both trees and interval graphs 
(more details are given in Section \ref{section-chordal}).  
We give additional
results showing how to compute upper bounds on $\beta$ for more
general classes of graphs in Sections \ref{section-bigger-beta}
and \ref{section-examples}.

Among these tools for bounding $\beta$, 
one of particular interest to our hypothetical combinatorial
auctioneer is the following generalization of the fact that
intersection graphs of subtrees have $\beta$ equal to one.  Suppose
that we have an \buzz{object graph} whose nodes are objects and in
which an edge exists between any two objects that are relevant to each
other in some way.  (In the car example, there might be an edge
between a wheel and its axle but not between a wheel and the hood
ornament.)  We demand that the objects in each bid be \buzz{germane}
in the sense that they must form a connected node subset of the object
graph.  
For many sparse object graphs, the intersection graph of all
connected sets of vertices
can be ordered so that a later set intersects an earlier set
only if it intersects a ``frontier set'' that may be much smaller than
the earlier set.  It is immediate that $\beta$ for the intersection
graph is bounded by the size of the largest frontier set
(more details are given in
Lemma~\ref{lemma-intersection-graph}).
Examples of such graphs are those of low treewidth 
(Theorem~\ref{theorem-treewidth}) and planar graphs 
(Corollary~\ref{corollary-planar}).

In Section \ref{section-budget} we show how to handle
more complex constraints on acceptable sets of bids.  
We investigate scenarios where
bids are grouped by bidder, and that each bidder is
limited to some maximum number of winning bids (an \buzz{unweighted
budget constraint}), or some maximum total cost of winning bids (a
\buzz{weighted budget constraint}).  By charging later bids an
approximate opportunity cost for earlier bids in the same budget
groups, we can solve these problems approximately with ratio $\beta+1$
with unweighted constraints and $2\beta+3$ for weighted constraints.
The results for unweighted budget constraints can be further
generalized for more complicated constraints.

Finally, in Section~\ref{section-open} we discuss some open problems
suggested by the current work.

\section{Simple combinatorial auctions}
\label{section-noBudgetConstraints}

In this section, we describe our algorithms for approximating the
maximum weight independent set problem, 
the opportunity cost algorithm
and
the local ratio opportunity cost algorithm .
Both algorithms return the same
approximation.

\subsection{The opportunity cost algorithm}
\label{subsection-opportunityCost}

We will write $u \pred v$ if $uv \in E$ and
call $u$ a \buzz{predecessor} of $v$ and $v$ a \buzz{successor} of $u$.  
The set of all predecessors of $u$ will be written as $\delta^-(u)$
and the set of all successors as $\delta^+(u)$.

Given a directed acyclic graph $G_0 = (V_0,E_0)$ with weights $\wt(v)$ for
each $v$ in $V$, the opportunity cost algorithm, \opcost{,} proceeds in two stages:

\begin{enumerate}
\item[\oc1]
Traversing the nodes according to the topological order of $G_0$,
compute a \emph{value} $\val(u)$ for each node $u$.
This value represents an estimate of the
gain we expect by including $u$ in the independent set; it is computed
by taking the weight of $u$ and subtracting off an \emph{opportunity
cost} consisting of the values of earlier positive-value nodes 
that conflict with $u$.  Formally, let
\begin{equation}
\val(u) = \wt(u) - \sum_{v \pred u} \max(0, \val(v)).
\label{eqn-value} 
\end{equation}

\item[\oc2]
Processing the nodes in reverse topological order, add any node
with non-negative value to the desired
independent set $\bids$ and discard its predecessors.
Formally, let
\begin{equation}
\select(u) = [\val(u) \ge 0] \wedge \forall v \in \delta^{+}(u):
 \neg \select(v).
\label{eqn-select}
\end{equation}
\end{enumerate}

The output of the algorithm is the set $\bids$ defined as 
all $u$ for which
$\select(u)$ is true. 
This set $\bids$ is clearly independent.
In Section \ref{subsection-approxRatioNoConstraints}, 
we examine how close $\bids$ is to optimal.

\subsection{The local ratio opportunity cost algorithm}
\label{subsection-localRatio}

The \emph{local ratio technique} can be used to recursively
find approximate solutions 
to
optimization problems over vectors in $\R^n$, subject to a set of
feasibility constraints. It was originally developed by Bar-Yehuda and
Even~\cite{BE:localratiotheorem}, and later extended by Bafna \etal{}
\cite{BBF:localratio}, Bar-Yehuda~\cite{BarYehuda:localratio}, and
Bar-Noy \etal{}~\cite{BarNoyetal2000}.

Let $w\in\R^n$ be a \buzz{weight} vector.  Let $F$ be a set of feasibility
constraints. A vector $x\in\R^n$ is a \buzz{feasible solution} to a
given problem $(F,w)$ if it satisfies all the constraints in $F$. The
\buzz{$w$-weight} of a feasible solution $x$ is defined to be the dot-product
$w\cdot x$; for $r \ge 1$, $x$ is an \buzz{$r$-approximation} with
respect to $(F, w)$ if $r \cdot w\cdot x \ge w\cdot x^*$, where $x^*$
is a feasible solution maximizing the $w$-weight. An algorithm is said
to have an \buzz{approximation ratio} of $r$ if it always returns an
$r$-approximate solution.


\begin{lemma}[Local Ratio Lemma \cite{BE:localratiotheorem}]
\label{lemma-local-ratio}
Let $F$ be a set of feasibility
constraints. Let $w$, $w_1$ and $w_2$ be weight vectors such that
$w = w_1 + w_2$. If $x$ is an $r$-approximation with respect to
$(F, w_1)$ and $(F, w_2)$, then $x$ is an $r$-approximation with
respect to $(F, w)$.
\end{lemma}

We now describe the local ratio opportunity cost algorithm, \lropcost{.} 
Given a directed acyclic graph $G_0 = (V_0, E_0)$ with
weights $\wt(v)$ for each $v\in V_0$, we pass $(G_0, \wt(\cdot))$ to the following recursive procedure. This procedure takes as input a graph $G$ and a weight function $w$ and proceeds as follows:
\begin{enumerate}
\item[\lr1] Delete all nodes in $G$ with non-positive weight. Let this new graph be $G_2$.
\item[\lr2] If $G_2$ has no nodes, return the empty set.
\item[\lr3] Otherwise, select a node $u$ with no predecessors 
in $G_2$,
and decompose the weight function $w$ as
$w = w_1 + w_2$, where
$$
w_1(v) = 
\begin{cases}
w(u) & \text{if $v \in \{u\}\cup \delta^+(u)$,} \\
0 & \text{otherwise,}
\end{cases}
$$
and $w_2 = w - w_1$.
\item[\lr4] Solve the problem recursively using $(G_2, w_2)$ as input. Let $\bids_2$ be the approximation to a maximum weight independent set returned by this recursive call.
\item[\lr5] If $\bids_2 \cup \{u\}$ is an independent set, return $\bids = \bids_2
\cup \{u\}$. Otherwise, return $\bids = \bids_2$.
\end{enumerate}

\begin{theorem}
\label{theorem-same}
\opcost{} and \lropcost{} return the same approximation to a maximum weight independent set.
\end{theorem}
\begin{proof}
Consider a recursive call $C$ of \lropcost{.} Let $u$ be the node that is selected to be processed in step \lr3. All of $u$'s predecessors in the original graph $G_0$ have either been processed in a previous step \lr3 or deleted in some step \lr1. 
Therefore,
the current weight of $u$, $w(u)$, as seen by the recursive call $C$, is just $\val(u)$, as defined in step
\oc1 of \opcost{.} Furthermore, we add node $u$ to our independent set in step \oc2 if and only if we add $u$ to our independent set in step \lr5.
\end{proof}

\subsection{Approximation ratios} 
\label{subsection-approxRatioNoConstraints}

\begin{theorem}
\label{theorem-approxRatioNoConstraints}
\opcost{} and \lropcost{} return a $\beta(G)$-approximation to a maximum weight independent set. Furthermore, there exist weights for which this bound is tight.
\end{theorem}
\maybe{20}{The proof is given in the full paper.  The essential idea
is to show by induction that $\bids_2$ is
$\beta$-approximation with respect to $w_2$, 
from which it follows that $\bids_2 \cup \{u\}$ is also a
$\beta$-approximation since $w_2(u) = 0$.
Showing that whichever we choose is a $\beta$-approximation with
respect to $w_1$ requires analyzing only the behavior of each in the
immediate directed neighborhood of $u$.  The full result is obtained
by applying Lemma \ref{lemma-local-ratio}.
}{
\begin{proof} We will prove the result for \lropcost{.} The full result follows from Theorem~\ref{theorem-same}. Clearly, the returned set of nodes $\bids$
is an independent set. By Lemma~\ref{lemma-local-ratio}, we need only show
that $\bids$ is a $\beta$-approximation with respect to $w_1$ and
$w_2$. We will prove this by induction on the recursion. The base case
of the recursion is trivial, since there are no positive weight nodes.

For the inductive step, assume that $\bids_2$ is a
$\beta$-approximation with respect to $w_2$. Then $\bids$ is also a
$\beta$-approximation with respect to $w_2$ since $w_2(u) = 0$ and
$\bids \subset \bids_2 \cup \{u\}$.

To show that $\bids$ is a $\beta$-approximation with respect to $w_1$,
we will derive an upper bound $\beta w(u)$ on the maximum $w_1$-weight
independent set and a lower bound $w(u)$ on the $w_1$-weight of any
\emph{$u$-maximal} independent set of nodes. 
A $u$-maximal independent
set of nodes either contains $u$ or adding $u$ to it violates the
property that it is an independent set. Our $w_1$ performance bound
is $\beta w(u) / w(u) = \beta$. 
Note that only $u$ and its successor nodes will have a
nonzero contribution to $w_1$-weight. 

The total weight of a maximum $w_1$-weight independent set 
is at most
$\beta(u)w(u) \le \beta(G) w(u) = \beta w(u)$. 
The total weight of any 
$u$-maximal independent set 
is at least $w(u)$,
since any such set
contains at least one element of $u \cup \delta^+(u)$, and all such
nodes are assigned weight $w(u)$.
Since the algorithm always chooses a $u$-maximal set, its
$w_1$ performance bound is $\beta$.

To show the bound is tight, pick some $v$ that maximizes
$\beta(v)$, and assign it weight $1$ and all of its successors
weight $1-\epsilon$, where $\epsilon > 0$.
Let every other node in $G$ have weight $0$.
When we run \opcost{}, the value of $v$ will be $1$, the
value of each of its successors will be
$-\epsilon$, and the value of any other node is irrelevant because it
has zero weight.  Thus \opcost{} returns a set of total
weight $1$ but the maximum weight independent set has total weight at least
$\beta(u) \cdot (1-\epsilon)$.
\end{proof}
}

\maybe{20}{}{
\subsection{Running time} 
\label{subsection-runningTimeNoConstraints}

\begin{theorem}
\label{theorem-running-time}
The running times of both \opcost{} and \lropcost{} are linear in the
size of the input graph $G_0$. 
\end{theorem}
\begin{proof}
\opcost{} computes $\val(v)$ for each node $v$
in time proportional to its indegree,
and computes $\select(v)$ for each node 
in time proportional to its outdegree,
for a total time of $O(|V_0|+|E_0|)$. 
In the case of \lropcost{,} a recursive call is made at most once for
each node in the graph, and defining $w_1$ and $w_2$ in each call
takes time proportional to the node's outdegree, for a total running
time of $O(|V_0| + |E_0|)$.
\end{proof}
}

\section{Properties of $\beta$}

For any $v$, $\beta(v)$ is at most 
the larger of $1$ or the outdegree of $v$.
Thus,
$\beta(G)$ is at most the larger of $1$ or the maximum
degree of $G$. 
In many cases we can use the structure of $G$ to get a
much better bound.

\subsection{Graphs with $\beta = 1$}
\label{section-chordal}

Graphs with orientations for which 
$\beta = 1$ can be characterized completely.
These are the \buzz{chordal} graphs, also known as \buzz{triangulated} graphs or
\buzz{rigid circuit} graphs.  
The defining property of a chordal graph is that no cycle of
length $4$ or more appears as an induced subgraph.  
A succinct discussion of these graphs, including a variety of
characterizations as well as several examples of interesting families
of chordal graphs,
can be found in \cite[pp.~280--281]{GLS}.
For our purposes the most useful of these characterizations are
stated in the following lemma:
\begin{lemma}
\label{lemma-chordal}
Let $G$ be an undirected graph.  Then the following properties of $G$ are
equivalent:
\begin{enumerate}
\item $G$ is chordal.
\item $G$ is the intersection graph of subtrees of a forest.
\item $G$ has an ordering $G'$ for which the successors of any node form a
clique.  Such an ordering is called a \buzz{perfect elimination
ordering}.  Restated in terms of $\beta$, $G$ has an ordering $G'$
for which $\beta(G') = 1$.
\end{enumerate}
\end{lemma}
\begin{proof}
See \cite[pp.~280-281]{GLS}.
\end{proof}

Chordal graphs can be recognized and ordered using a
specialized version of breadth-first search in
$O(|V|+|E|)$ time as shown by Rose \etal{} \cite{RoseTL76}, and their maximum
cardinality independent sets can be computed in $O(|V|+|E|)$ time
as shown by Gavril
\cite{Gavril72}.  Gavril's algorithm is essentially the same as 
step \oc1
of the opportunity cost algorithm; it chooses all
nodes with positive value and works because the sets $\{v : u \pred
v\}$ for each $u$ in the independent set form a clique covering.  
However, this algorithm does not deal with weights.

Special cases of graphs with $\beta=1$
include trees, interval graphs, and disjoint unions of
cliques.  The last are particularly nice: 

\begin{lemma}
\label{lemma-cliques}
Let $G$ be a disjoint union of cliques.  Then \emph{every} orientation
$G'$ of $G$ has $\beta(G') = 1$.
\end{lemma}
\begin{proof}
For each $u$ in $G'$, $\delta^+(u)$ is a clique.
\end{proof}

\subsection{Graphs with larger $\beta$ values}
\label{section-bigger-beta}

For general graphs, we cannot compute $\beta$ even approximately.
However, we can bound the $\beta$ values of many graphs using the
tools in this section.

\begin{lemma}
\label{lemma-unions-and-subgraphs}
Let $G$ be a directed graph.
\begin{enumerate}
\item 
If $G = G_1 \cup G_2$,
then $\beta(G) \le \beta(G_1) + \beta(G_2)$.
\item 
If $G$ is a node-induced subgraph of $H$, then $\beta(G) \le \beta(H)$.
\end{enumerate}
\end{lemma}
\begin{proof}
Let $u$ be a node of $G$.  Let $\delta^+(u)$, $\delta^+_1(u)$, 
$\delta^+_2(u)$, and $\delta^+_H(u)$ 
be the set of all successors of $u$ in $G$,
$G_1$, and $G_2$, respectively.  Let $A$ be any independent
subset of $\delta^+(u)$.  Then
\begin{enumerate}
\item 
$|A| 
\le 
|A \cap \delta^+_1(u)|
+
|A \cap \delta^+_2(u)|
\le
\beta(G_1)
+ \beta(G_2)$, and
\item
$A$ is an independent subset of $\delta^+_H(u)$, implying $|A| \le
\beta(H)$.
\end{enumerate}
\end{proof}

\suppress{
\begin{lemma}
\label{lemma-union}
Let $G = G_1 \cup G_2$.
Then $\beta(G) \le \beta(G_1) + \beta(G_2)$.
\end{lemma}
\maybe{90}{}{
\begin{proof}
Consider some node $u \in G$.  Let $\delta^+(u)$, $\delta^+_1(u)$, and
$\delta^+_2(u)$ be the set of all successors of $u$ in $G$,
$G_1$, and $G_2$, respectively.  Let $A$ be a maximum independent
set in $\delta^+(u)$.  Then
$|A| 
\le 
|A \cap \delta^+_1(u)|
+
|A \cap \delta^+_2(u)|
\le
\beta(G_1)
+ \beta(G_2).$
\end{proof}
}

\begin{lemma}
\label{lemma-subgraph}
If $G$ is a node-induced subgraph of $H$, then $\beta(G) \le \beta(H)$.
\end{lemma}
\maybe{90}{}{
\begin{proof}
Consider some node $u \in G$.  Let $A$ be a maximal independent
subset of the set $\delta^+_G(u)$ of successors of $u$ in $G$.  Then
$A$ is also an independent subset of the set $\delta^+_H(u)$ of
successors of $u$ in $H$.
\end{proof}
}
}

\maybe{110}{}{
The following technical lemma will be useful for bounding $\beta$ for
intersection graphs.
}

\begin{lemma}
\label{lemma-intersection-graph}
Let $G$ be the intersection graph of a set system $\allbids$ whose
union is $\objs$.  
Let $G$ be ordered by an ordering $<$ such that for each 
$A \in \allbids{}$ there exists a ``frontier set'' $S_A \subseteq U$
of size at most $k$,
so that if
$A < B$ and $A \cap B \ne \emptyset$, then
$S_A \cap B \ne \emptyset$.
Then $\beta(G) \le k$.
(Note that $S_A$ need not be contained in $A$.)
\end{lemma}
\maybe{60}{}{
\begin{proof}
Let $B_1, \ldots, B_l$ be some independent set of successors of $A$.
Under the conditions of the lemma each $B_i$ intersects $S_A$.  But
since the $B_i$ do not themselves intersect, each must intersect $S_A$
in a distinct element.  Thus there are at most $k$ of them.
\end{proof}
}

\maybe{60}{The converse of Lemma \ref{lemma-intersection-graph}
does not hold, although we
observe in the full paper that a result
similar to the converse holds if independence
number is replaced by clique covering number in the definition of
$\beta$.
}{
The converse of the lemma does not hold.  
Instead, its proof
shows that the \buzz{clique covering number} $\overline{\chi}$ of
$\delta^+(A)$ 
(defined as
the minimum size of any set of cliques whose union is $\delta^+(A)$)
is at most $k$, since the set of all $B$ that intersect
$S_A$ at any particular element form a clique.  
Note that \emph{any} directed acyclic graph in which 
$\overline{\chi}(\delta^+(v))$ is bounded can be represented as an
intersection graph with small frontier sets as in
Lemma~\ref{lemma-intersection-graph},\footnote{The trick is to add a new common element to all
members of
each clique, and let $S_A$ be the set of all such
new elements for the cliques that cover $\delta^+(A)$.}
in general the
independence number of $\delta^+(v)$
may be smaller than the clique
covering number.
}

When $\allbids{}$ consists of
connected node subsets of some graph $H$, we can obtain good orderings of
the intersection graph $G$ of $\allbids{}$ by exploiting the structure of $H$.  
\suppress{
In particular, we can
generalize one direction of Lemma~\ref{lemma-chordal}, by showing that
$H$ itself has low $\beta(H)$ whenever it has low treewidth.}

We start by reviewing the definition of treewidth.
A \buzz{tree decomposition} of an \emph{undirected}
graph $H = (V,E)$ consists of a tree $T$ and a
family of sets ${\cal V} = \{V_t\}$ where $t$ ranges over nodes of $T$,
satisfying the following three properties:
\begin{enumerate}
\item $\bigcup_{t\in T} V_t = V$.
\item For every edge $uv$ in $E$, there is some $V_t$ that contains
both $u$ and $v$.
\item If $t_2$ lies on the unique path from $t_1$ to $t_3$ in $T$,
then $V_{t_1} \cap V_{t_3} \subseteq V_{t_2}$.
\end{enumerate}

The \buzz{width} of a tree decomposition $(T, {\cal V})$ is $\max
|V_t|-1$.  The \buzz{treewidth} $\tw(H)$ of a graph $H$ is the
smallest width of any tree decomposition of $H$.

\begin{theorem}
\label{theorem-treewidth}
If $G$ is an intersection graph of connected node subsets $\allbids{}$
of some graph
$H$ with treewidth $k$, then there is an orientation $G'$ of $G$ with
$\beta(G') \le k+1$.
Given $\allbids{} = \{A_i\}$ and a tree decomposition 
$(T, {\cal V} = \{V_t\})$
of $H$, this orientation can
be computed in time $O(\sum_i |A_i| + |T| + \sum_t |V_t|)$, which is
linear in the size of the input.
\end{theorem}
\maybe{100}{
\maybe{10}{}{The proof is given in the full paper.
The intuition is that with an appropriate ordering
we can 
get frontier sets for Lemma \ref{lemma-intersection-graph} from
cuts in the tree.}}{
\begin{proof}
Let $(T, {\cal V})$ be a tree decomposition of $H$ with width $k$.  We
will use this tree decomposition to construct an ordering of the
connected node subsets of $H$, with the property that if $A < B$ then
either $A \cap B = \emptyset$ or $B$ intersects some frontier
set $S_A$ with at most $k+1$ elements.  
The full result then follows from Lemma \ref{lemma-intersection-graph}.

Choose an arbitrary root $r$ for $T$, and let $t_1 \ge t_2$ if
$t_1$ is an ancestor of $t_2$ in the resulting rooted tree.
Extend the resulting partial order to an arbitrary linear order.
For each connected node subset $A$ of $H$, let
$t_A$ be the greatest node in $T$ for which $V_{t_A}$ intersects $A$.
Given two connected node subsets $A$ and $B$ of $H$, let $A < B$ if $t_A <
t_B$ and extend the resulting partial order to any linear order.

Ordering $T$ can be done in $O(|T|)$ time using depth first search.
We can then compute and the maximum node in $T$ containing each node
of $H$ in time $O(\sum_t V_t)$ by considering each $V_t$ in order.
The final step of ordering the $A_i$ in the given set system ${\cal
S}$ takes $O(\sum_i |A_i|)$ time, since we must examine each element
of each $A_i$ to find the maximum one.  The total running time is thus
linear in the size of the input.

Now suppose $A \le B$ in this ordering.  We will show that any such $B$
intersects $V_{t_A}$, and thus that $V_{t_A}$ is our desired frontier set
$S_A$.
There are two cases.

If $t_A = t_B$, we are done.

The case $t_A < t_B$ is more complicated.
We will make heavy use of a lemma from \cite{RobertsonS86}, which
concern the effect of removing some node $t$ from $T$.  
Their Lemma 2.3 implies that if
$x,x'$ are not in $V_t$, then either $x$ and $x'$ are separated in $H$
by $V_t$ or $x$ and $x'$ are in the same branch (connected component)
of $T-t$.

Let $p$ be the parent of $t_A$ (which exists because $t_A$ is not
the greatest element in the tree ordering).  
We have $A \cap V_p = \emptyset$ since $p > t_A$.
Since $A$ is a connected set, it cannot be separated without removing
any of its nodes; thus by Lemma 2.3 every element of $A$ is in the
same branch of $T-p$, which consists precisely of the subtree of $T$
rooted at $t_A$.

Now $B$ contains at least one node $x$ in
the vertex set of an element of
the subtree rooted at $t_A$, and at least one node $x'$ in $V_{t_B}$,
which is not in this subtree because $t_B > t_A$.
So by Lemma 2.3 of \cite{RobertsonS86}, either one of $x,x'$ is in
$V_{t_A}$ or $B$ is separated by $V_{t_A}$.  In the latter case $B$
intersects $V_{t_A}$ since $B$ is also connected.
\end{proof}
}

Applying Theorem~\ref{theorem-treewidth} to planar graphs gives:

\begin{corollary}
\label{corollary-planar}
If $G$ is the intersection graph of a family  $\allbids{}$ of
connected node subsets
of a planar graph $H$ with $n$ nodes,
then there is an orientation $G'$ of $G$ with
$\beta(G') = O(\sqrt{n})$.
Given $H$,
a data structure of size $O(n)$ can be precomputed in
time $O(n \log n)$
that allows this orientation $G'$ to be computed for
any
$\allbids{} = \{A_i\}$ 
in time $O(\sum_i |A_i|)$.
\suppress{
If $G$ is an intersection graph of connected node subsets of a planar
graph $H$ with $n$ nodes, then there is an orientation $G'$ of $G$ 
with $\beta = O(\sqrt{n})$.}
\end{corollary}
\begin{proof}
Reed~\cite{Reed92} gives a recursive $O(n \log n)$ 
algorithm for computing tree
decompositions of constant-treewidth graphs based on a linear time
algorithm for finding approximate separators for small node subsets.
Replacing this separator-finding subroutine with the linear time algorithm
of Lipton and Tarjan~\cite{LiptonT79} gives 
an $O(n \log n)$ time algorithm for computing a tree decomposition of
a planar graph.  Since each separator has size at most $k = O(\sqrt{n})$, the
resulting tree decomposition has width at most $4k = O(\sqrt{n})$ by
Theorem~1 of \cite{Reed92}.

Since all we need to compute a good ordering of $\allbids$ is the
ordering of the $n$ nodes, we can compute this ordering as described
in the proof of Theorem~\ref{theorem-treewidth} and represent it in
$O(n)$ space by assigning each node an index in the range $1$ to $n$.
Ordering $\allbids$ then takes linear time as described in the proof
of Theorem~\ref{theorem-treewidth}.
\end{proof}

\subsection{Examples}
\label{section-examples}

Applying the results of Sections~\ref{section-chordal} and
\ref{section-bigger-beta} gives:

\begin{enumerate}
\item A linear-time
algorithm for finding a maximum weight independent set
of an interval graph, since $\beta(G) =
1$ by Lemma \ref{lemma-chordal}, and since chordal graphs can be
recognized and ordered in linear time using the work of
Rose \etal{}~\cite{RoseTL76}.

While the maximum independent set problem is easily solved for this
case (for example, by using 
the linear time interval graph recognition 
algorithm of Hsu and Ma~\cite{HM1999}
followed by a simple application of dynamic programming)
this is an
example of how our general method yields good algorithms as special
cases.
\item As another
special case, a 2-approximation algorithm for
interval selection of Berman and DasGupta\cite{BermanDasGupta}.
Here intervals are
partitioned into groups and we must choose non-overlapping intervals
with at most one per group. 
The bid graph $G$ is of the form 
$G_1 \cup G_2$
where
$G_1$
is an interval graph
and
$G_2$
is a disjoint union of cliques, one for each group.
Thus $\beta(G)=2$ by
Lemmas \ref{lemma-chordal}, \ref{lemma-cliques}, and
\ref{lemma-unions-and-subgraphs}.
\item
A 3-approximation algorithm for ``double auction'' interval selection
where each interval has both a seller and a buyer, and at most one
interval per seller or buyer may be selected.  This is the same as the
previous case except the graph is now $G_1 \cup G_2 \cup G_3$
where
$G_2$ and $G_3$ are both disjoint unions of cliques.
\item
In general, a mechanism for taking \emph{any} bid graph with
$\beta = k$ and adding up to $m$ such unique-selection constraints
to get a $(k+m)$-approximation algorithm by 
repeated applications
of
Lemmas \ref{lemma-cliques} and \ref{lemma-unions-and-subgraphs}.
So for example we get a 3-approximation algorithm for
maximum weight three-dimensional matching and a 4-approximation
algorithm for auctioning off tracts of undeveloped
land spanning intervals where each 
tract must be
acceptable to a seller who provides it, a builder who will develop it, and a buyer who will ultimately purchase both the land and the buildings developed
on it.
\item
An algorithm to $k$-approximate a maximum weight independent set of
any subgraph of a $k$-dimensional rectangular grid.
Orient each edge to leave the point whose
coordinates have a smaller sum, giving $\beta \le k$.
\item
A linear-time algorithm for $2$-approximating a maximum weight
independent set of the intersection graph
of intervals on a cycle.  
This follows from Lemma \ref{lemma-intersection-graph}: order
connected node subsets by inclusion, extend to a linear order $\prec$, 
and observe
that if $A \prec B$ and $A$ intersects $B$ then $B$ intersects one of
$A$'s two endpoints.\footnote{One 
can do better by breaking the cycle to reduce it to a standard
interval graph problem (see, for example, the approach taken by
\cite{BarNoyetal2000}), but the $2$-approximation shows how one can
still do reasonably well with our general algorithms \opcost{} and
\lropcost{}.}
\item An algorithm for 
intersection graphs of bounded-height rectangles in a discrete
2D
grid.  
Order the rectangles by their largest $x$-coordinate, and make the
rightmost grid points of each rectangle be its frontier set in the
sense of Lemma \ref{lemma-intersection-graph}.  If each rectangle is at most
$h$ tall, there are at most $h$ grid points in each frontier.
This generalizes in the obvious way to higher
dimensions given bounds on all but one of the coordinates, in which
case
the approximation ratio becomes the product of the bounds.
\end{enumerate}

\subsection{Hardness of computing $\beta$}
\label{section-hardness}

The difficulty of even approximating the independence
number of a graph extends to the directed local independence number.

\begin{theorem}
\label{theorem-np-hard}
Any algorithm that can approximate $\beta(G)$ for an $n$-node
directed
acyclic graph $G$ with a ratio of
$f(n)$ can be used to approximate the size $\alpha(H)$ of a maximum
independent set of an undirected $n$-node graph $H$ with ratio
$f(n+1)$.
Thus by H{\aa}stad's bound on approximating a maximum clique
\cite{hastad1999.clique}, we cannot approximate $\beta$
by $O(n^{1-\epsilon})$ unless $P=NP$.
\end{theorem}
\maybe{80}{}{
\begin{proof}
Given an undirected $n$-node graph $H$, construct an $(n+1)$-node 
directed acyclic graph
$G$ by (a) directing
the edges of $H$ in any consistent order, and (b) adding a new
source node $s$ to $H$ with edges from $s$ to every node in $H$.

Let $I$ be an independent set in $H$.  Then every node in $I$ is a
successor of $s$ in $G$, and furthermore these nodes are all
independent.  It follows that $\beta(G) \ge \beta(s) \ge \alpha(H)$.

Conversely, if $I'$ is an independent set of successors of some node
$v$ in $H$, it cannot contain $s$ (since $s$ is not a successor of any
node), and thus $I'$ is also an independent set in $H$.  So we have
$\alpha(H) \ge \beta(G)$.
\end{proof}
}

\subsection{Effects of node ordering}
\label{section-ordering}


The performance of the opportunity cost
algorithm is strongly sensitive to the order in
which the nodes are processed, as this affects the value of
$\beta(u)$ for each node $u$.  For many of the examples given in
the Section \ref{section-examples},
a good ordering is provided by the structure of
the problem.  But what happens in a general graph?
\maybe{10}{In the full paper, we show (a) with the best ordering,
the opportunity cost algorithm is optimal, and (b) with the worst
ordering, the algorithm produces the same answer as the greedy
algorithm.}{

\begin{theorem}
\label{theorem-order-opt}
For any graph $G$ with given weights, 
there exists an orientation $G'$ of $G$ for which
both \opcost{} and \lropcost{} 
output a maximum independent set of $G$.
\end{theorem}
\maybe{60}{}{
\begin{proof}
Let $A$ be any independent set in $G$.  Choose
the ordering so that all nodes in $A$ precede all nodes not in $A$.
Then for any $u \in A$, $u$ has no predecessors in the oriented graph
and $\val(u) = \wt(u)$.  

Let $A'$ be the independent set computed by the algorithm.
If $u$ is in $A$ but not $A'$, it must have a successor $v$ in $A'-A$ 
with non-negative value.
Since the value of each $v$ is its weight less the
weight of all its
neighbors in $A$,
the total weight of all elements
of $A'-A$ must exceed the total weight of all elements in $A-A'$,
and we have 
$\wt(A') = \wt(A'-A) + \wt(A' \cap A) \geq \wt(A-A') + \wt(A' \cap A)
= \wt(A)$.
\end{proof}
}

\maybe{50}{}{
In a sense what Theorem~\ref{theorem-order-opt} 
shows is that finding a good ordering of a
general graph is equivalent to solving the
maximum weight independent set problem.
This is not surprising since evaluating $\beta(u)$ for even a
single node $u$ requires solving this problem.  It follows that to
get small approximation ratios we really do need to exploit some
special property of
the given graph.

In the other direction, we can show that there exist orderings that are
not very good:
}

\begin{theorem}
\label{theorem-order-greedy}
If all nodes in a graph $G$ have distinct weights, orienting $G$ in
order of decreasing weight causes \opcost{} and \lropcost{}
to return the same independent set as 
the greedy algorithm.
\end{theorem}
\maybe{80}{}{
\begin{proof}
We will prove the result for \opcost{}; by Theorem~\ref{theorem-same}
the same result holds for \lropcost{}.

Let $\pi$ order the nodes in order of
decreasing weight.  Let us show by induction on $\pi$ that if
the greedy algorithm chooses a node $v$, then $\val(v) = \wt(v)$; but
if the greedy algorithm does not choose $v$, then $\val(v) < 0$.
Suppose we are processing some node $v$ and that this induction
hypothesis holds for all nodes previously processed.  If the greedy
algorithm picks $v$, then all $v$'s predecessors were not chosen and
have negative value, and $\val(v) = \wt(v)$.  
If the greedy algorithm does not pick $v$, it is because 
it chose some $u \pred v$; now $\val(v) \le \wt(v) - \val(u) = \wt(v)
- \wt(u) < 0$.

Since the only nodes with non-negative weights are those chosen by the
greedy algorithm, \opcost{} selects them as its output.
\end{proof}
}
}

\section{Auctions with budget constraints}
\label{section-budget}

Consider the following bidding scenarios:

\begin{enumerate}
\item A bidder whose car has broken down wants to buy 
either a new engine, a new car, or an umbrella and a taxi ride
home, but doesn't particularly care which.  However, she has no
interest in winning more than one of these bids.

\item Another bidder wants to buy at most three 1968 Volkswagen
Beetle hood ornaments, but she would like to bid on all that are available
so as not to miss any.

\item Yet another bidder has only \$100 in cash, but would like to
place multiple bids totaling more than \$100, with the understanding that she
can only win bids up to her budget.
\end{enumerate}

All of these are examples of \buzz{budget constraints}, in which bids in some group consume a common scarce resource.  We would like to extend our algorithms to handle such constraints, which are natural in real-world bidding situations.

The first scenario is an example of a \buzz{1-of-$n$} constraint, where
at most one of a set of $n$ bids can be accepted.  
This special case can be handled by
modifying $G$ by forming a clique out of all bids in each set $S_i$;
under the assumption that the $S_i$ are disjoint, this increases
$\beta$ by at most $1$ (using Lemmas \ref{lemma-cliques} and
\ref{lemma-unions-and-subgraphs}).  
The second scenario depicts a more general \buzz{$k$-of-$n$} constraint.  
Such constraints are handled by extending our algorithms
to account for the possible revenue loss from bids that cannot be
selected because the budget constraint has been exceeded.  
Again, the approximation ratio rises by $1$.
We refer to both 1-of-$n$ and $k$-of-$n$ constraints as
\buzz{unweighted budget constraints}, as each bid consumes a single
unit of the budget.

\buzz{Weighted budget constraints}, exemplified by the third scenario, are
more complicated. With such constraints, we must ensure that the sum
of the weights of accepted bids in some group $S$ is at most some
bound $b$.  
A complication arises because 
a maximal allowed set of bids might only
fill half of a budget limit.  With some additional
modifications to our algorithms, we get a performance bound of
$2\beta+3$.

\subsection{Unweighted budget constraints}
\label{section-unweighted}

Suppose the bids are partitioned into groups $S_1,\ldots,S_r$ and that
no more than $k_i$ bids may be selected from $S_i$, for $1\le i\le
r$. For each bid $u$, let $g(u)$ denote the index of the group to
which $u$ belongs and let $S_u = S_{g(u)}$ and $k_u = k_{g(u)}$.

\unweighted{\opcost{}} is an extension of \opcost{} to handle unweighted budget constraints. It has a similar two-step structure. 

In the first step, like \oc1, we traverse the nodes in topological order and compute a value for each node. We must extend the definition of value for each node to account for the \emph{possible} revenue loss from previously processed bids that may not be selected in the second step because of the budget constraint:
\begin{equation}
\label{eqn-unweighted-value}
\val(u)
 = \wt(u) - \sum_{v\rightarrow u}\max(0, \val(v)) - 
 \frac{1}{k_u}\cdot\sum_{v\in S_u-\{u\}, v < u} \max(0, \val(v)),
\end{equation}
where the notation $v < u$ means that $v$ has already been processed
(before $u$). Note that the inclusion of $u$ in the set of winning bids
does not necessarily preclude previously processed bids in $S_u$ from
also being selected---they may also be selected if the budget $k_u$
allows.  The coefficient $\frac{1}{k_u}$ scales the opportunity cost to account for this fact. 

In the second step, like \oc2, we traverse the bid graph in
reverse topological order,
selecting nodes of positive value whose addition to those already
selected does not violate the independence or budget constraints.

\maybe{50}{We define the unweighted-budget-constraint local ratio
opportunity cost algorithm similarly (more details are given in the
full paper.)}{
\unweighted{\lropcost{}} solves the same problem using the local ratio technique. 
It follows the same structure as \lropcost{.} We begin by deleting all non-positive weight nodes from the graph. If any nodes remain, we select a node $u$ with no predecessors, and decompose the weight function into $w = w_1 + w_2$. This time, the decomposition must account for bids that are in the same budget group. We define
$$ w_1(v) = 
\begin{cases}
w(u) & \text{if $v\in \{u\} \cup \delta^+(u)$,} \\
\frac{1}{k_u} w(u) &\text{if $v \in S_u - \{u\}$,} \\
0 & \text{otherwise,}
\end{cases}
$$
and recursively solve the problem using $w_2$ as the weight function. After the recursive call, we must decide if we should add $u$ to the set of winning bids $\bids_2$. In \lropcost{,} we added $u$ to $\bids_2$ if and only if $\bids_2 \cup \{u\}$ was an independent set. In this algorithm, we must also ensure that the budget constraints are satisfied before adding $u$ to $\bids_2$. We say that a set of bids is \emph{feasible} if they form an independent set and the budget constraints are satisfied.
}

\begin{theorem}
\label{theorem-unweighted}
Given a directed bid graph $G$, a partition of the nodes of $G$ into
nonempty subsets
$S_1,\ldots,S_r$, and an unweighted budget constraint $k_i$ for each $S_i$,
\begin{enumerate}
\item 
\unweighted{\opcost{}} and \unweighted{\lropcost{}} return the same
approximation to a revenue maximizing set of bids.
\item
\unweighted{\opcost{}} and \unweighted{\lropcost{}} $(\beta(G) + 1)$-approximate
an optimal set of bids.
\item
\unweighted{\opcost{}} and \unweighted{\lropcost{}}
run in time linear in the size of $G$.
\end{enumerate}
\end{theorem}
\begin{proof}
The proof that both algorithms return the same approximation is
similar to the proof of Theorem~\ref{theorem-same}.

The proof of the approximation ratio follows the same structure as
the proof of Theorem~\ref{theorem-approxRatioNoConstraints}. We prove
the result for \unweighted{\lropcost{.}} By 
Lemma~\ref{lemma-local-ratio}, we
need only show that the returned set of bids $\bids$ is a $(\beta +
1)$-approximation with respect to $w_2$ and $w_1$. We do this using
induction on the recursion. The fact that $\bids$ is a $(\beta +
1)$-approximation with respect to $w_2$ follows trivially from the
inductive assumption.

In the case of $w_1$, we will derive an upper bound $U$ on the maximum $w_1$-weight of a set of feasible bids and a lower bound $L$ on the $w_1$-weight of any \emph{$u$-maximal} set of bids. A $u$-maximal set of bids either contains $u$ or adding $u$ to it would violate the feasibility constraints. In the case of a set of feasible bids, its total $w_1$-weight is at most
$\beta(u)w(u) + \frac{k_u}{k_u} w(u) \le w(u)(\beta + 1) = U$, since the only nonzero contribution to $w_1$-weight comes from $\delta^+(u)$ and $S_u$. In the case of a $u$-maximal set of bids, if $u$ cannot be added to the set, then either (1) a successor of $u$ is already in the set, in which case the total $w_1$-weight is at least $w(u)$, or (2) the budget constraint is exceeded, in which case the total $w_1$-weight is at least $w(u)$. Therefore, the $w_1$-weight of these bids is at least $w(u)$ and the $w_1$ performance bound is
$$
\frac{U}{L} = \frac{w(u)(\beta + 1)}{w(u)} = \beta + 1.
$$

The proof of the running time is similar to the proof of
Theorem~\ref{theorem-running-time}.
All of the steps that \unweighted{\opcost{}} and
\unweighted{\lropcost{}}
share with \opcost{} and \lropcost{} take linear time.  
\unweighted{\opcost{}} adds the cost of computing the last term in
(\ref{eqn-unweighted-value}).  
Storing
$\sum_{v \in S_i} \max(0, \val(v))$ in a variable $\Delta_{S_i}$
for each $S_i$ allows this term
to be computed in time $O(1)$ for each node, with an additional $O(1)$
cost per node to update the appropriate $S_i$.
The same technique allows budget constraints to be tested in $O(1)$
time per node during the second step.
Thus the additional time is linear.

The corresponding modification to \unweighted{\lropcost{}} similarly adds
only linear time.
Rather than updating the weight of each node $v$
before each recursive call, we will compute the ``current'' weight of
each node $v$ as it is required, subtracting off the total weight
$\Delta_{S_v}$ of all
previously-processed nodes in $S_v$ as in \unweighted\opcost{}. 
\end{proof}

\maybe{20}{The above analysis assumes that the budget constraints
partition the bids.  For some applications (e.g., bids involving
matching up buyers with sellers), we may have overlapping constraints.
Overlapping constraints may also be used to handle bids for identical
items in limited supply, by
grouping all bids asking for copies of the same item together.

In the full paper we give an extended version of the unweighted budget
algorithm, and show:
\begin{theorem}
\label{theorem-overlapping-unweighted}
Given a bid graph $G$ and a set of unweighted budget constraints with
maximum overlap $t$, the extended unweighted budget algorithm
approximates the optimal set of bids within $\beta(G) + t$.
\end{theorem}
}{
\subsection{Overlapping unweighted constraints}
\label{section-overlapping}

The analysis in Section~\ref{section-unweighted}
assumes that the budget constraints
partition the bids.  For some applications (e.g., bids involving
matching up buyers with sellers), we may have overlapping constraints.
Overlapping constraints may also be used to handle bids for identical
items in limited supply, by
grouping all bids asking for copies of the same item together. The algorithms described above can be generalized to handle overlapping constraints.

Suppose we have a family of $r$ sets of bids
$\mathcal{S} = \{S_1,\ldots, S_r\}$,
that each bid appears in at most $t$ of these sets, and
that at most $k_i$ bids may be accepted from set $S_i$.

In \overlapping{\unweighted{\opcost{,}}} when computing the value of a node $u$, we need to account for the possible revenue loss from nodes in each set that $u$ belongs to:
$$
\val(u) 
= \wt(u) 
  - \sum_{v\rightarrow u}\max(0, \val(v))  
  - 
\sum_{
1\le i \le r,
u \in S_i
}
\left(
\frac{1}{k^i}\sum_{
v\in S_u,
v < u
}\max(0, \val(v))\right).
$$
The rest of the algorithm is the same as \unweighted{\opcost{.}} 

In \overlapping{\unweighted{\lropcost{,}}} the only change from \unweighted{\lropcost{}} is in the decomposition of the weight function. We decompose it as
$$ w_1(v) = 
\begin{cases}
w(u) & \text{if $v\in \{u\} \cup \delta^+(u)$,} \\
\sum_{1\le i\le r, u,v\in S_i}\frac{1}{k_i} w(u) &\text{if there exist $S_i$ containing both $u$ and $v$,} \\
0 &\text{otherwise.}
\end{cases}
$$

\begin{theorem}
\label{theorem-overlapping-unweighted}
Given a directed bid graph $G = (V,E)$, a family of nonempty node subsets 
$S_1,\ldots,S_r$, where each node appears in at most $t$ of the $S_i$,
and an unweighted budget constraint $k_i$ for each $S_i$,
\begin{enumerate}
\item 
\overlapping{\unweighted{\opcost{}}} and
\overlapping{\unweighted{\lropcost{}}} return the same
approximation to a revenue maximizing set of bids.
\item
\begin{sloppypar}
\overlapping{\unweighted{\opcost{}}} and
\overlapping{\unweighted{\lropcost{}}}
\mbox{$(\beta(G) + t)$-approximate}
an optimal set of bids.
\end{sloppypar}
\item
\overlapping{\unweighted{\opcost{}}} and
\overlapping{\unweighted{\lropcost{}}}
run in time $O(|V|t + |E|)$
\end{enumerate}
\end{theorem}
\begin{proof}
Similar to the proof of Theorem~\ref{theorem-unweighted}.
The additional $O(|V|t)$ term comes from having to apply up to $t$
budget constraints to each node; since $\sum_{i} |S_i| \leq |V|t$,
this term also covers the cost of reading the $S_i$ from the input and
initializing the variables for each subset.
\end{proof}
}

\subsection{Weighted budget constraints}

Suppose that bids are partitioned into groups $S_1,\ldots,S_r$ and
that the total value of the winning bids from group $i$ can be no more
than $b_i$. For each bid $u$, let $g(u)$ denote the index of the group
to which $u$ belongs and let $S_u = S_{g(u)}$ and $b_u = b_{g(u)}$.

This case is more complicated than the unweighted case. The difficulty
arises when estimating a lower bound on the $w_1$-weight of a
$u$-maximal set of bids $S$. If $u$ cannot be added to the set because
the budget constraint will be exceeded, the $w_1$-weight of $S$ can be
as small as $\epsilon$, if $w_1(u) = b_u$.

\maybe{20}{

We solve this problem by running the algorithm separately for
``heavy'' bids with $w(v) > \frac{1}{2}b_v$ and ``light'' bids with
$w(v) \le \frac{1}{2}b_v$.  Because a bidder can win at most one heavy
bid, we can approximate the best choice of such bids within $\beta+1$
using the unweighted-budget algorithm of
Section~\ref{section-unweighted}.
For light bids, we use a variant of the algorithm where 
the $\frac{1}{k_u}$ multiplier is replaced by $\frac{2}{b_u}$, which
we show gives a $\beta+2$ approximation.  We then take the better
solution, to get a $2\beta+3$ approximation overall.

}{

We will describe changes required to \lropcost{} to
handle this case.  Corresponding changes can be made to \opcost{.}
We will run variations of the algorithm twice, once for the \emph{heavy}
bids $v$ with $w(v) > \frac{1}{2}b_v$ and once for the \emph{light}
bids $v$ with $w(v) \le \frac{1}{2}b_v$. We then return the
better of the two solutions. 

In \heavy\weighted\lropcost{,} we put an unweighted budget constraint of 1 on each bidder and run \unweighted\lropcost{.}

\begin{lemma}
\label{lemma-heavy}
\heavy\weighted\lropcost{} $(\beta+1)$-approximates an optimal 
set of
heavy bids.
\end{lemma}
\begin{proof}
Since each heavy bid consumes more than half a bidder's budget, each
bidder can win at most one bid. This is just a simple unweighted
budget constraint and can be solved as described in Section
\ref{section-unweighted} for a performance bound of $\beta + 1$.
\end{proof}

In \light\weighted\lropcost{,} when decomposing the weight function, we set
$$
w_1(v) = 
\begin{cases}
w(u) & \text{if $v \in \{u\} \cup \delta^+(u)$,} \\
\frac{2}{b_u} w(v) w(u) &\text{if $v \in S_u - \{u\}$,} \\
0 & \text{otherwise.}
\end{cases}
$$
\maybe{50}{In the full paper, we show that the light-bid algorithm
approximates the best light-bid solution within $\beta+2$.
}{Before adding $u$ to the winning set of bids $\bids_2$, we must ensure that it does not conflict with other bids in $\bids_2$ and that the weighted budget constraint is not violated. The rest of the algorithm is identical to \lropcost{.}
\begin{lemma}
\label{lemma-light}
\light\weighted\lropcost{} $(\beta+2)$-approximates an optimal set of
light bids.
\end{lemma}
\begin{proof}
This proof uses the same structure and notation as the proof of
Theorem~\ref{theorem-unweighted}.
An upper bound $U$ on the $w_1$-revenue of any feasible set of bids is
$w(u)(\beta  + 2)$. With regards to a $u$-maximal set of
bids, if $u$ cannot be added to the set because the budget constraint
$b_u$ will be exceeded, the existing bids in the set must have weight
at least $b_u/2$, since $w(u) \le b_u/2$. A lower bound $L$ on the
$w_1$-revenue of any $u$-maximal set of bids is therefore $w(u)$. The performance bound of this algorithm is $\frac{U}{L} = \beta + 2$, as claimed.
\end{proof}
}
\begin{theorem}
\label{theorem-weighted}
Given a directed bid graph $G$, a partition of the nodes of $G$ into
nonempty subsets
$S_1,\ldots,S_r$, and a weighted budget constraint $b_i$ for each $S_i$,
\begin{enumerate}
\item 
\weighted{\opcost{}} and \weighted{\lropcost{}} return the same
approximation to a revenue maximizing set of bids.
\item
\weighted{\opcost{}} and \weighted{\lropcost{}} 
$(2\beta(G) + 3)$-approximate an optimal set of bids.
\item
\weighted{\opcost{}} and \weighted{\lropcost{}}
run in time linear in the size of $G$.
\end{enumerate}
\end{theorem}
\begin{proof}
The sum of the optimal revenues for the heavy and light bids is at
least equal to the optimum revenue among all bids. From
Lemmas~\ref{lemma-heavy} and ~\ref{lemma-light}, the better of the two
solutions will be within a factor of $2\beta + 3$ of the optimum for
the general problem.

For the running time, observe that decomposing the bids into heavy and
light bids takes linear time, that \heavy\weighted\opcost{}
and \heavy\weighted\lropcost{} are equivalent to
\unweighted\opcost{} and \unweighted\lropcost{} and thus take linear
time by Theorem~\ref{theorem-unweighted},
and that 
\light\weighted\opcost{}
and
\light\weighted\lropcost{}
can be made to run in linear time using techniques similar to those
used for \unweighted\opcost{} and \unweighted\lropcost{}.
\end{proof}
}

\section{Further Research}
\label{section-open}
This paper opens up several directions for further research.  An
immediate open problem is whether overlapping weighted budget
constraints can be processed as efficiently as their unweighted
counterparts are processed in
Theorem~\ref{theorem-overlapping-unweighted}.

It would be of importance to compare the performance of our algorithms
and others in practice.  The comparison could be conducted on
simulations, but it would be more useful to analyze the performance on
real auction data.

As the examples of car sales and land sales demonstrate,
topological structures exist in actual bids.  Another good example is
the FCC auction of airwaves in 1994 and 1995
\cite{MM1996}, where each trading area is an auction object, the
trading areas form a plane graph, and bidders prefer to acquire
contiguous trading areas. It would be useful to examine past auctions
to determine whether similar connectivity structures exist and how
such structures affect the computational complexity of bidding
strategies and winner determination.

\bibliographystyle{siam}
\bibliography{opcost}

\end{document}